# Teleporting digital images

**Mario Mastriani**

http://orcid.org/0000-0002-5627-3935

**Abstract:** During the last 25 years the scientific community has coexisted with the most fascinating protocol due to Quantum Physics: quantum teleportation (QTele), which would have been impossible if quantum entanglement, so questioned by Einstein, did not exist. In this work, a complete architecture for the teleportation of Computational Basis States (CBS) is presented. Such CBS will represent each of the possible 24 classical bits commonly used to encode every pixel of a 3-color-channel-image (red-green-blue, or cyan-yellow-magenta). For this purpose, a couple of interfaces: classical-to-quantum (Cl2Qu) and quantum-to-classical (Qu2Cl) are presented with two versions of the teleportation protocol: standard and simplified.

---

## 1. Introduction

### 1.1 Quantum Teleportation

In 1993 a fundamental paper for the history of Physics was published [1], where a protocol for the teleportation of an unknown quantum state was proposed without violating the No-Cloning Theorem [2]. This protocol uses a strange phenomenon of Quantum Mechanics (QMech) known as Quantum Entanglement (QEnta), in which two or more particles leave their individualities to become part of something unique and integral. These particles retain the aforementioned attribute even if they are separated to opposite places in the universe.

Now, if we measure one of those particles, the entanglement disappears, the joint wave function collapses, and if the measured particle becomes a spin-up, the other instantaneously becomes a spin-down, and vice-versa. This instantaneity is given independently of the distance that mediates between both particles, contradicting the very essence of Special Relativity (SRela) [3], which postulates that nothing can travel faster than light. This curious result was called by Einstein as "spooky action at a distance", and constitutes the center of one of the most famous paradoxes in the history of Physics suggested by Albert Einstein, Boris Podolsky and Nathan Rosen known as EPR paradox [4]. There is a palliative to this controversy which consists in the impossibility of transmitting useful information through an instantaneous link based on entanglement [1]. This palliative constitutes a hot border between two of the main pillars of Physics: the SRela [3] and QMech [5].

In 1964 John Bell [6] tried to solve this controversy by proposing a theorem based on an inequality by which the non-locality of QMech could be established as long as it violated such inequality. Experiments carried out by Aspec [7-9] and others (with and without loop-holes) apparently established the non-locality of QMech, contradicting what was manifested by the EPR paradox. Part of the scientific community does not question the non-locality of QMech, since there is an absolute consensus, however, they are hesitant about the quality of the mentioned experiments.

Since 1997, a series of experiments have demonstrated the practical feasibility of QTele protocol (QTele) [10-18], even in the presence of noise [19-22]. Thirteen years later, Prof. Hotta [23] demonstrated how energy could be teleported, triggering all kinds of speculations about the possible teleportation of matter, by virtue of the close link between matter and energy starting from the famous Einstein equation, $E = m\ c^2$, for a particle at rest. Currently, a simplified version of the original QTele protocol [24, 25] opens up a whole new range of possibilities in Quantum Communications (QComm) [26-29] thus completing the arsenal of essential tools for Quantum Technology to be used in the future.

*1.2 Superdense Coding*

A complementary protocol to the QTele, called Superdense Coding (SDC) [1], allows us to send classical bits through a quantum channel. Besides, SDC is the foundational basis of a couple of interfaces needed when handling CBS in a QComm context [27]. Recently, a new SDC protocol based on a simplified version of QTele has come out with remarkable results [25]. In fact, the mentioned protocol is presented as the most probable basis for secure communication between Earth and Mars, in a future mission to Mars.

*1.3 Interfaces*

The need for Cl2Qu and Qu2Cl interfaces in all branches of Quantum Information Processing (QIP) is evident [30], however, in no other case as in Quantum Image Processing (QImP) has it become so evident. Works like Quantum Boolean Image Denoising [31] highlight the imperative need for an efficient interface between each of the possible 24 classical bits commonly used to encode each pixel of an image with 3 color channels (red, green, blue) and the internal representation that said classical bits must have within the quantum computer, that is, as CBS. This conversion cannot result from a qubits preparation procedure because the images are too large, this must be done automatically thanks to an interface. For example, for an image of 1920 columns, 1080 rows, 8 bits-by-color, and 3 colors (an image of common size these days), we must prepare 49,766,400 qubits (in this particular case: CBS). This is highly impractical, since, how long would a similar amount of qubit preparations take us in a laboratory? The problem of Quantum Measurement (QuMe) [32] has confined the practicality of QImP to the exclusive use of CBS, as it is exposed in [33]. If we wanted to work in QImP with generic qubits, we would find ourselves with a problem that does not exist to date even with Cl2Qu interface for this type of qubit. Besides, it is impossible to recover exactly the Quantum Algorithm (QAlgo) result because of QuMe [32], as explained in [33]. It happens that the measurement noise due to QuMe is greater than that admissible in a standard process of Digital Image Processing (DIP) [34-37]. The distortion in the recovery of a qubit at the output of a QAlgo is greater than that normally accepted [33], being the Heisenberg Uncertainty Principle (HUP) and the complementarity [30], the exclusive responsible for it, as explained in [33]. This is the problem that fundamentally affects the internal representation technique known as Flexible Representation of Quantum Images (FRQI) [38] and all its variants. Besides, another famous technique within QImP is that known as Novel Enhanced Quantum Representation (NEQR) [39]. The problem with NEQR, as well as all its variants, is that it is not a Cl2Qu Interface, however, they need one. In fact, if we had a Cl2Qu interface, then why would we need NEQR? On the other hand, the reason why both techniques (FRQI and NEQR) are accepted within QImP resides in the fact that all the papers that mention them only involve implementations in a high-level interpreter such as MATLAB® [40], and not on an optical table as it should be. Under these circumstances everything seems to work, but when done on an optical table the real outcome is very different.

**2. Materials and Methods**

*2.1 A brief of Quantum Information Processing*

A generic quantum bit (or qubit) can be expressed thanks to the superposition principle [30] as,

$$|\psi\rangle = \alpha|0\rangle + \beta|1\rangle, \tag{1}$$

where $|\psi\rangle$ is called *wave function*, $\alpha$ and $\beta$ are complex numbers such that $|\alpha|^2 + |\beta|^2 = 1$, and the states $|0\rangle = \begin{bmatrix} 1 & 0 \end{bmatrix}^T$ and $|1\rangle = \begin{bmatrix} 0 & 1 \end{bmatrix}^T$ (being $[\bullet]^T$ the transpose of $[\bullet]$) are understood as different polarization states of light allocated at the poles of the Bloch's sphere (Figure 1).

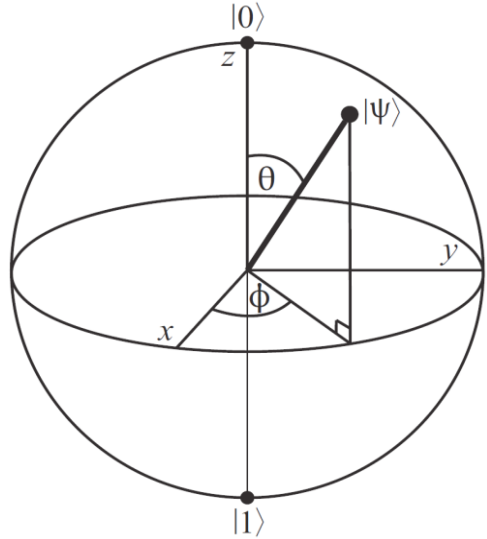

**Figure 1.** Bloch's Sphere.

These states $\{|0\rangle, |1\rangle\}$ constitute an orthonormal basis and they are known as CBS and can be expressed in several ways:

$$Spin\ up = |\uparrow\rangle = |0\rangle = \begin{bmatrix} 1 \\ 0 \end{bmatrix} = 0\ qubit\ basis\ state = North\ Pole \tag{2}$$

and

$$Spin\ down = |\downarrow\rangle = |1\rangle = \begin{bmatrix} 0 \\ 1 \end{bmatrix} = 1\ qubit\ basis\ state = South\ Pole \tag{3}$$

On the other hand, a more general analysis of the wave function $|\psi\rangle$ allows us to establish that

$$|\psi\rangle = e^{i\gamma}\left( cos\frac{\theta}{2}|0\rangle + e^{i\phi} sin\frac{\theta}{2}|1\rangle \right) = e^{i\gamma}\left( cos\frac{\theta}{2}|0\rangle + \left( cos\phi + i\ sin\phi \right) sin\frac{\theta}{2}|1\rangle \right) \tag{4}$$

where $0 \le \theta \le \pi$, $0 \le \phi < 2\pi$. We can ignore $e^{i\gamma}$ from Equation (4), because it has no observable effects [30], therefore, we can effectively write:

$$|\psi\rangle = cos\frac{\theta}{2}|0\rangle + e^{i\phi} sin\frac{\theta}{2}|1\rangle \tag{5}$$

The angles $\theta$ and $\phi$ define a point on the unit three-dimensional sphere (Figure 1), with,

$$\alpha = cos\frac{\theta}{2}, \quad \beta = e^{i\phi}\, sin\frac{\theta}{2} \tag{6}$$

Except in the case where $|\psi\rangle$ is one of the CBS $|0\rangle$ or $|1\rangle$, in all other cases, the representation needs to use the superposition principle [30]. The parameters $\theta$ and $\phi$ can be re-interpreted as spherical coordinates specifying a point

$$\vec{a} = \left(a_x, a_y, a_z\right) = \left(sin\,\theta\, cos\,\phi\,,\, sin\,\theta\, sin\,\phi\,,\, cos\,\theta\right) \tag{7}$$

on the unit sphere in $\Re^3$ (according to Equation 4). In other words, the state of a qubit is a unit vector in the two-dimensional complex Hilbert's space $\mathbb{C}^2$. Finally, a column vector $|\psi\rangle$ is called a *ket* vector $\begin{bmatrix} \alpha\ \beta \end{bmatrix}^{\mathrm{T}}$, while a row vector $\langle\psi|$ is called a *bra* vector $\begin{bmatrix} \alpha^*\ \beta^* \end{bmatrix}$, where, $(\bullet)^*$ means complex conjugate of $(\bullet)$.

*2.2 Color decomposition and bit slicing*

First, we need to decompose all digital images in their 3 color components (red, green, blue) [31]. Thus, we will obtain 24 bitplanes (8 bitplanes for every color) thanks to a procedure known as bit slicing [31]. We get 8 bitplanes per color, where the 7th bitplane (the closest to the observer) is called the Most Significant Bit (MSB) being the most morphologically committed bitplane with the original image [31] (Figure 2). Conversely, bitplane 0 (the furthest from the observer) is the Least Significant Bit (LSB) and the least morphologically committed bitplane with the original image. Below, we provide the MATLAB® code [40] necessary for slicing:

```
function Ibpp = slicer(I,bpp)
% Casting of the algorithm:
% bpp = bit-per-pixel
% I = Each color component of the image
% Ibpp = I in bpp bitplanes (strictly binary)
[ROW,COL] = size(I);
for r = 1:ROW
  for c = 1:COL
    aux = d2b(I(r,c)-1,bpp);
    for b = 1:bpp
      Ibpp(r,c,b) = aux(b);
    end
  end
end
return;
```

```
function bvpp = d2b(p,bpp)
% Casting of algorithm:
% d = bit depth
% p = pixel value
% bvpp = binary vector per pixel
bvpp = zeros(1,bpp);
d = 1;
while p > 0,
  bvpp(d) = mod(p,2);
  p = p/2;
  p = floor(p);
  d = d+1;
end
bvpp = rot90(rot90(bvpp));
return;
```

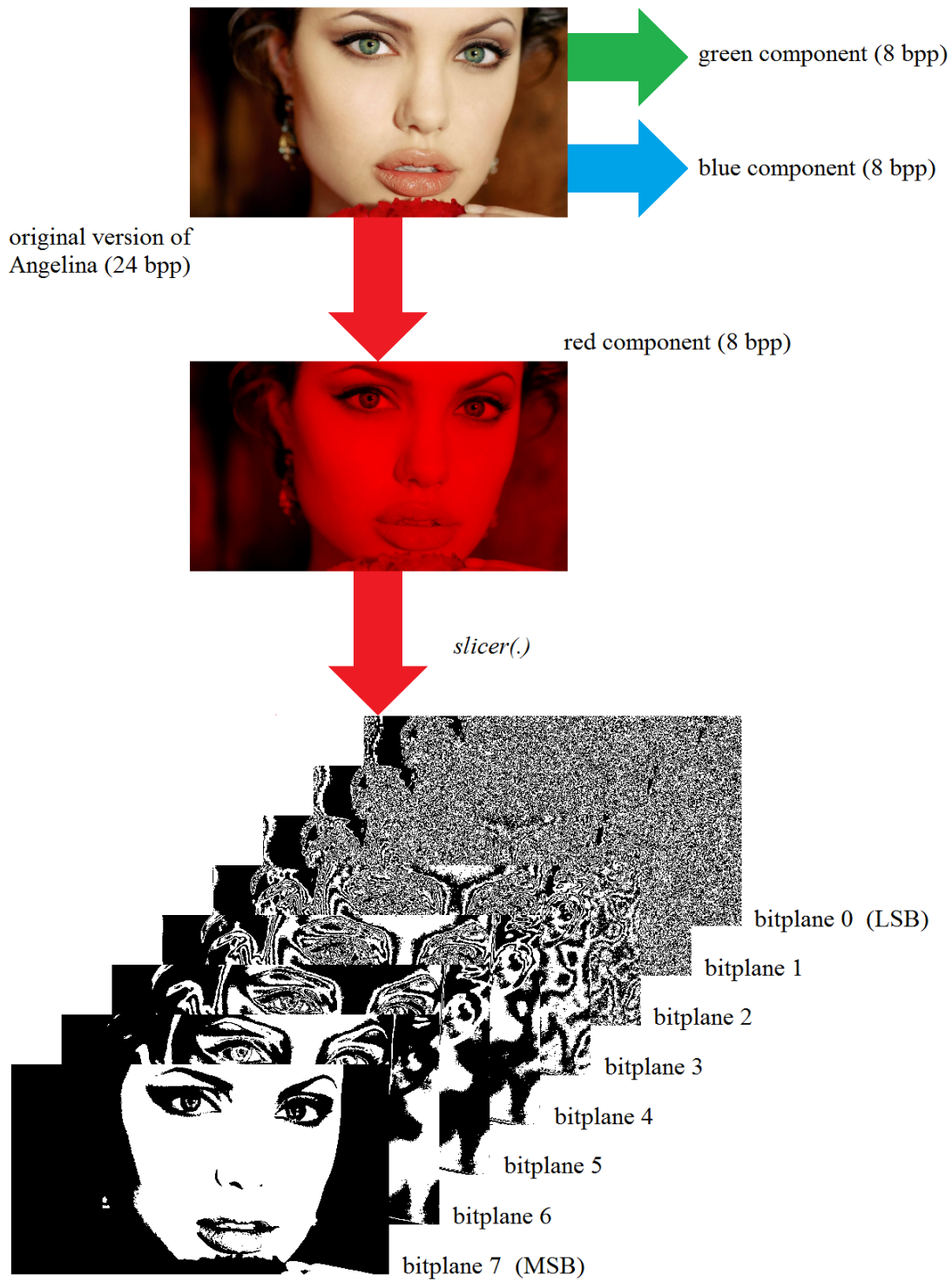

green component (8 bpp)

blue component (8 bpp)

original version of
Angelina (24 bpp)

red component (8 bpp)

slicer(.)

bitplane 0  (LSB)
bitplane 1
bitplane 2
bitplane 3
bitplane 4
bitplane 5
bitplane 6
bitplane 7  (MSB)

**Figure 2.** Bitplanes of the red component for Angelina obtained by slicing, with special remarks for MSB and LSB.

It is evident, from here on, that an element in black in every bitplane is a classical bit equal to 1 and will be represented with a qubit equal to $|1\rangle$, while, an element in white in every bitplane is a classical bit equal to 0 and will be represented with a qubit equal to $|0\rangle$ in a future Cl2Qu interface.

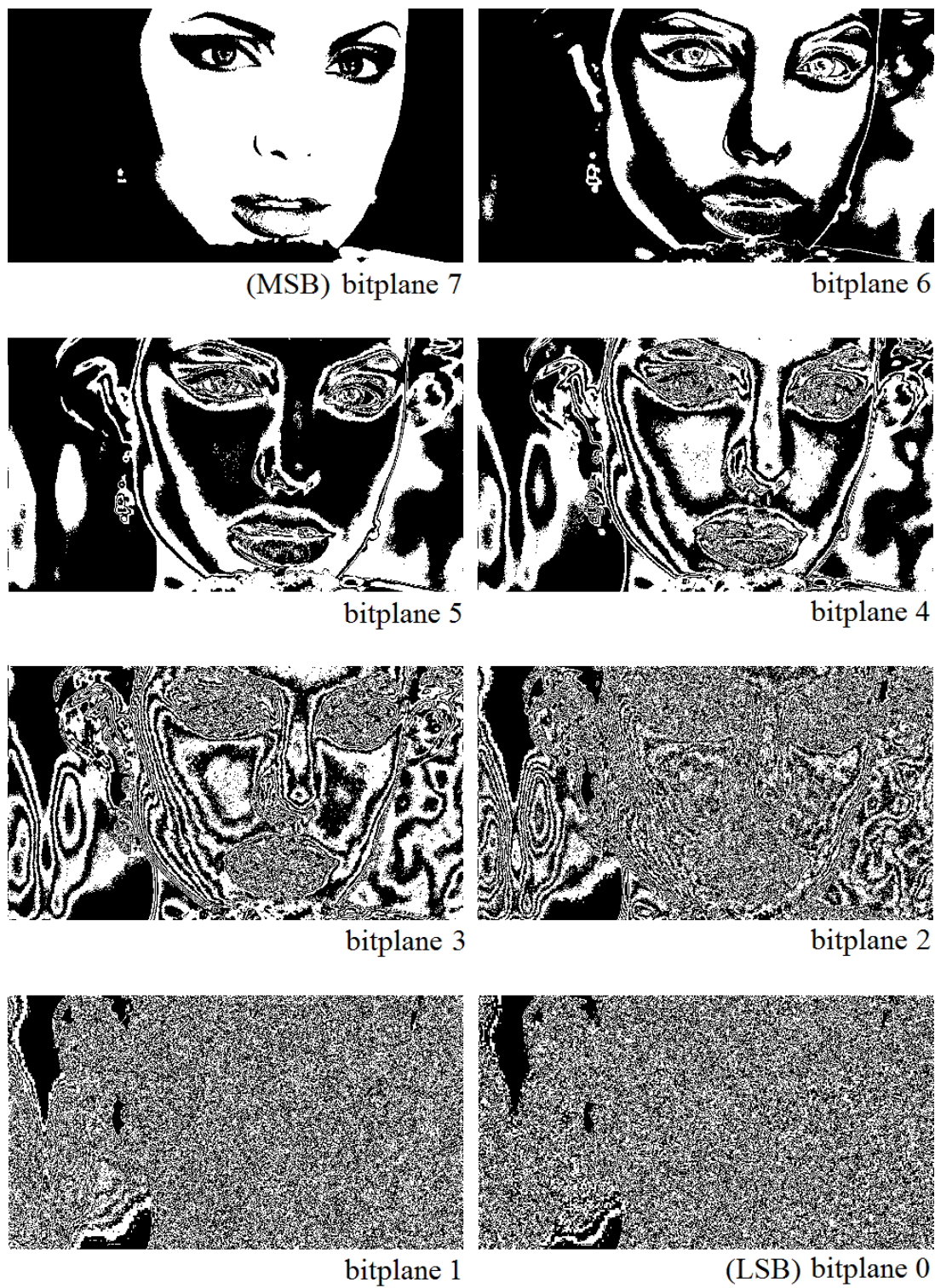

**Figure 3.** Angelina and her 8 bitplanes, including MSB and LSB.

Figure 3 shows the 8 bitplanes of Angelina for the red channel in detail, from MSB (bitplane 7) to LSB (bitplane 0). Let us observe that as we move from MSB to LSB, different bitplanes are increasingly unrecognizable compared to the original image, i.e., Angelina. As we can see, LSB is completely different from the original morphology of Angelina's picture. This is one reason why the LSB is Steganography territory [41]. The other reason is that any change in the LSB does not produce visually detectable changes in the original image.

*2.3 Standard Quantum Teleportation - noiseless analysis*

QTele begins with the distribution of an EPR pair between Alice and Bob. We can choose any of the EPR of the complete set of Bell's basis:

$$\left|\beta_{00}\right\rangle = \left|\Phi^{+}\right\rangle = \frac{1}{\sqrt{2}}\left(\left|00\right\rangle + \left|11\right\rangle\right), \quad \left|\beta_{01}\right\rangle = \left|\Phi^{-}\right\rangle = \frac{1}{\sqrt{2}}\left(\left|00\right\rangle - \left|11\right\rangle\right)$$

$$\left|\beta_{10}\right\rangle = \left|\Psi^{+}\right\rangle = \frac{1}{\sqrt{2}}\left(\left|01\right\rangle + \left|10\right\rangle\right), \quad \left|\beta_{11}\right\rangle = \left|\Psi^{-}\right\rangle = \frac{1}{\sqrt{2}}\left(\left|01\right\rangle - \left|10\right\rangle\right)$$

(8)

We normally choose $\left|\beta_{00}\right\rangle$. This distribution constitutes the entanglement link between Alice and Bob. After that, we continue with the complete sketch of QTele of Figure 4, where the green line indicates the border between Alice's and Bob's sides, that is, both extremes of the entanglement link. In Figure 4, a single fine line represents a wire carrying one qubit, while a double line represents a wire carrying one classical bit [30], while, the classical channel is really a control classical channel for disambiguation purposes (as we will see below through two bits), while the entanglement link is really an entanglement data link. On the other hand, in Figure 4, the following blocks mean: *SPD* (single photon detectors), {$\sigma_x$, $\sigma_z$} are Pauli's matrices activated by the bits {$b_2$, $b_1$} respectively [1, 30], and EPR is the source of $\left|\beta_{00}\right\rangle \equiv \left|\Phi_{+}^{A\cup B}\right\rangle$ of Equation (8).

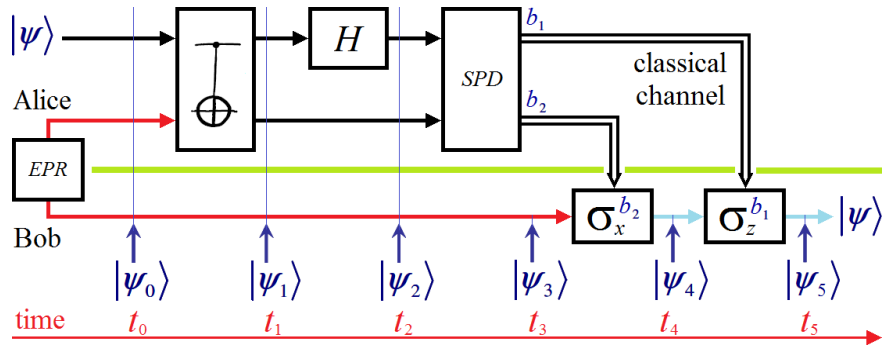

**Figure 4.** Standard teleportation protocol using an EPR pair and two classical bits for disambiguation.

Now, If $\left|\psi_{0}\right\rangle = \left|\psi\right\rangle = \alpha\left|0\right\rangle + \beta\left|1\right\rangle$ is an arbitrary and unknown state to be teleported with $\left|\alpha\right|^{2} + \left|\beta\right|^{2} = 1$ and $\alpha \wedge \beta \in \mathbb{C}$ of a Hilbert's space, then, the initial state (3-partite state) will be,

$$\left|\psi_{0}\right\rangle = \left|\psi\right\rangle \otimes \left|\beta_{00}\right\rangle = \left|\psi\right\rangle \left|\beta_{00}\right\rangle = \left(\alpha\left|0\right\rangle + \beta\left|1\right\rangle\right) \frac{1}{\sqrt{2}}\left(\left|00\right\rangle + \left|11\right\rangle\right)$$

$$= \frac{1}{\sqrt{2}}\left[\alpha\left|0\right\rangle\left(\left|00\right\rangle + \left|11\right\rangle\right) + \beta\left|1\right\rangle\left(\left|00\right\rangle + \left|11\right\rangle\right)\right] = \frac{1}{\sqrt{2}}\left[\alpha\left|000\right\rangle + \alpha\left|011\right\rangle + \beta\left|100\right\rangle + \beta\left|111\right\rangle\right]$$

$$= \left[\frac{\alpha}{\sqrt{2}} \quad \frac{\beta}{\sqrt{2}} \quad 0 \quad 0 \quad 0 \quad 0 \quad \frac{\alpha}{\sqrt{2}} \quad \frac{\beta}{\sqrt{2}}\right]^{T}$$

(9)

where for simplicity (and from here on) we have adopted $\left|x\right\rangle \otimes \left|y\right\rangle = \left|x\right\rangle\left|y\right\rangle$ in a generic form, and this operation is done inside a *beamsplitter*.

Now, a CNOT gate is applied to Equation (9),

$$\left|\psi_1\right\rangle = \frac{1}{\sqrt{2}}\left[\alpha\left|000\right\rangle + \alpha\left|011\right\rangle + \beta\left|110\right\rangle + \beta\left|101\right\rangle\right]$$

$$= \begin{bmatrix} \alpha\!\!\diagup\!\!\sqrt{2} & 0 & 0 & \beta\!\!\diagup\!\!\sqrt{2} & 0 & \beta\!\!\diagup\!\!\sqrt{2} & \alpha\!\!\diagup\!\!\sqrt{2} & 0 \end{bmatrix}^T$$

(10)

In practice, Kronecker's product and CNOT gate are implemented together on the same *beamsplitter* [10, 11, 16-18]. Then, we apply a Hadamard's gate to the elements of Equation (10),

$$\left|\psi_2\right\rangle = \frac{1}{2}\left[\left|00\right\rangle\sigma_x^0\sigma_z^0\left|\psi\right\rangle + \left|01\right\rangle\sigma_x^1\sigma_z^0\left|\psi\right\rangle + \left|10\right\rangle\sigma_x^0\sigma_z^1\left|\psi\right\rangle + \left|11\right\rangle\sigma_x^1\sigma_z^1\left|\psi\right\rangle\right]$$

$$= \frac{1}{2}\left[\left|\Phi^+\right\rangle\sigma_x^0\sigma_z^0\left|\psi\right\rangle + \left|\Phi^-\right\rangle\sigma_x^1\sigma_z^0\left|\psi\right\rangle + \left|\Psi^+\right\rangle\sigma_x^0\sigma_z^1\left|\psi\right\rangle + \left|\Psi^-\right\rangle\sigma_x^1\sigma_z^1\left|\psi\right\rangle\right]$$

$$= \begin{bmatrix} \alpha\!\!\diagup\!\!2 & \alpha\!\!\diagup\!\!2 & \beta\!\!\diagup\!\!2 & -\beta\!\!\diagup\!\!2 & \beta\!\!\diagup\!\!2 & -\beta\!\!\diagup\!\!2 & \alpha\!\!\diagup\!\!2 & \alpha\!\!\diagup\!\!2 \end{bmatrix}^T$$

$$= \begin{bmatrix} \alpha\!\!\diagup\!\!2 & 0 & \beta\!\!\diagup\!\!2 & 0 & 0 & 0 & 0 & 0 \end{bmatrix}^T \rightarrow \left|\Phi^+\right\rangle \rightarrow \left|00\right\rangle \rightarrow 00 \rightarrow \sigma_x^0\sigma_z^0$$

$$+ \begin{bmatrix} 0 & \alpha\!\!\diagup\!\!2 & 0 & -\beta\!\!\diagup\!\!2 & 0 & 0 & 0 & 0 \end{bmatrix}^T \rightarrow \left|\Psi^+\right\rangle \rightarrow \left|10\right\rangle \rightarrow 10 \rightarrow \sigma_x^0\sigma_z^1$$

$$+ \begin{bmatrix} 0 & 0 & 0 & 0 & \beta\!\!\diagup\!\!2 & 0 & \alpha\!\!\diagup\!\!2 & 0 \end{bmatrix}^T \rightarrow \left|\Phi^-\right\rangle \rightarrow \left|01\right\rangle \rightarrow 01 \rightarrow \sigma_x^1\sigma_z^0$$

$$+ \begin{bmatrix} 0 & 0 & 0 & 0 & 0 & -\beta\!\!\diagup\!\!2 & 0 & \alpha\!\!\diagup\!\!2 \end{bmatrix}^T \rightarrow \left|\Psi^-\right\rangle \rightarrow \left|11\right\rangle \rightarrow 11 \rightarrow \sigma_x^1\sigma_z^1$$

(11)

The last rows of Equation (11) represent Alice's options inside the *SPD*. Alice randomly selects one of the bases and performs the measurement, transmitting to Bob the corresponding classical bits through a classical channel. Alice's options within the SPD are equally probable and the random choice that she makes of the base has to do with being sure not to clone the original state between her and Bob [10, 11].

Table 1 synthesizes the complete process of QTele, where Alice measures two of the possible qubits of the basis of Equation (8), and therefore, she transmits the corresponding bits $b_1$ and $b_2$ via a classical channel to Bob. The QMeas process is imperative to make the wave-function of the original arbitrary state collapse since this is necessary to do so as not to violate the No-Cloning Theorem [2]. In other words, the QMeas process destroys the original arbitrary state [30, 32] eliminating any possibility of cloning.

**Table 1.** Alice's side: measurement of the base, classical transmission of bits, and the collapse of states, Bob's side: classical reception of bits, gates application for the final recovery of the arbitrary state.

| Alice's measurement | Alice transmits | This happens with probability | Collapsed state | Bob applies $\sigma_x^{b_1}\sigma_z^{b_2}$ |
|---|---|---|---|---|
| $\left|\Phi^+\right\rangle \rightarrow 00$ | $b_2 b_1 = 00$ | $\left\|\frac{1}{2}\sigma_x^0\sigma_z^0\left|\psi\right\rangle\right\|^2 = \frac{1}{4}$ | $\left|\Phi^+\right\rangle\sigma_x^0\sigma_z^0\left|\psi\right\rangle$ | $\sigma_x^0\sigma_z^0\left|\psi\right\rangle = \left|\psi\right\rangle$ |
| $\left|\Psi^+\right\rangle \rightarrow 01$ | $b_2 b_1 = 01$ | $\left\|\frac{1}{2}\sigma_x^1\sigma_z^0\left|\psi\right\rangle\right\|^2 = \frac{1}{4}$ | $\left|\Psi^+\right\rangle\sigma_x^1\sigma_z^0\left|\psi\right\rangle$ | $\sigma_x^1\sigma_z^0\left|\psi\right\rangle = \sigma_x\left|\psi\right\rangle$ |
| $\left|\Phi^-\right\rangle \rightarrow 10$ | $b_2 b_1 = 10$ | $\left\|\frac{1}{2}\sigma_x^0\sigma_z^1\left|\psi\right\rangle\right\|^2 = \frac{1}{4}$ | $\left|\Phi^-\right\rangle\sigma_x^0\sigma_z^1\left|\psi\right\rangle$ | $\sigma_x^0\sigma_z^1\left|\psi\right\rangle = \sigma_z\left|\psi\right\rangle$ |
| $\left|\Psi^-\right\rangle \rightarrow 11$ | $b_2 b_1 = 11$ | $\left\|\frac{1}{2}\sigma_x^1\sigma_z^1\left|\psi\right\rangle\right\|^2 = \frac{1}{4}$ | $\left|\Psi^-\right\rangle\sigma_x^1\sigma_z^1\left|\psi\right\rangle$ | $\sigma_x^1\sigma_z^1\left|\psi\right\rangle = \sigma_x\sigma_z\left|\psi\right\rangle$ |

The last column of Table 1 represents the local operations realized by Bob in $t_3$ and $t_4$ to reconstruct the original state $|\psi\rangle$. At this point, it is important to mention that in literature there are several concerns regarding the implementation of teleportation protocols using a bigger or smaller dimensional commitment but always with two classical bits for disambiguation. An interesting example can be found in [42], which shows that the one-qubit teleportation can be considered as a state transfer between subspaces of the whole Hilbert's space of an indivisible eight-dimensional system. However, this as well as the rest of the papers that manipulate high dimensional quantum systems for the implementation of QTele protocols do it with two classical bits for disambiguation, except in the case of the new protocol presented here which does not use disambiguation bits.

On Alice's side, the combination of the modules composed by the following gates: *CNOT*, *H* (Hadamard) and QMeas, constitute what is known as the Bell-State-Measurement (BSM), while on Bob's side, its modules are unitary operations necessary for the reconstruction of the teleported state. Alice's measurement and transmission of the classical bits of disambiguation along with Bob's unitary operations are the clearest examples of Local Operations and Classical Communications (LOCC) [43].

*2.4 Standard Quantum Teleportation - noisy analysis*

Starting again from Figure 4, and considering noise in the EPR pair by a disturbance of the shape

$$|\beta_{00}\rangle_n = A|00\rangle + B|11\rangle \tag{12}$$

where subscript *n* means *noise*, and

$$|A|^2 + |B|^2 = 1, \quad \text{with} \quad (A \neq B) \wedge \left(A \neq \frac{1}{\sqrt{2}}\right) \wedge \left(B \neq \frac{1}{\sqrt{2}}\right) \tag{13}$$

Then, repeating Equation (9) but with $|\beta_{00}\rangle_n$ instead of $|\beta_{00}\rangle$, we will have

$$
\begin{aligned}
|\psi_0\rangle = |\psi\rangle|\beta_{00}\rangle_n &= (\alpha|0\rangle + \beta|1\rangle)(A|00\rangle + B|11\rangle) \\
&= \alpha A|000\rangle + \beta A|100\rangle + \alpha B|011\rangle + \beta B|111\rangle
\end{aligned} \tag{14}
$$

Now, a CNOT gate is applied to Equation (14), resulting in

$$|\psi_1\rangle = \alpha A|000\rangle + \beta A|110\rangle + \alpha B|011\rangle + \beta B|101\rangle. \tag{15}$$

Then, we apply a Hadamard's gate to the elements of Equation (15),

$$
\begin{aligned}
|\psi_2\rangle &= \frac{1}{\sqrt{2}}\Big[\alpha A|000\rangle + \alpha A|100\rangle + \beta A|010\rangle - \beta A|110\rangle + \alpha B|011\rangle + \alpha B|111\rangle + \beta B|001\rangle - \beta B|101\rangle\Big] \\
&= \frac{A}{\sqrt{2}}|00\rangle\alpha|0\rangle + \frac{B}{\sqrt{2}}|00\rangle\beta|1\rangle + \frac{A}{\sqrt{2}}|10\rangle\alpha|0\rangle - \frac{B}{\sqrt{2}}|10\rangle\beta|1\rangle + \frac{A}{\sqrt{2}}|01\rangle\beta|0\rangle + \frac{B}{\sqrt{2}}|01\rangle\alpha|1\rangle + \frac{B}{\sqrt{2}}|11\rangle\alpha|1\rangle - \frac{A}{\sqrt{2}}|11\rangle\beta|0\rangle \\
&= \frac{1}{\sqrt{2}}\Big[|00\rangle(A\alpha|0\rangle + B\beta|1\rangle) + |10\rangle(A\alpha|0\rangle - B\beta|1\rangle) + |01\rangle(B\alpha|1\rangle + A\beta|0\rangle) + |11\rangle(B\alpha|1\rangle - A\beta|0\rangle)\Big]
\end{aligned} \tag{16}
$$

From here on, we will follow a procedure similar to that of Table 1 but taking into account how sensitively the state is affected by noise. In fact, seeing Equation (16), it is evident that it is almost impossible to recover the state $|\psi\rangle$ in an exact way [19-22, 24, 25].

### 2.5 Simplified Quantum Teleportation - noiseless analysis

Unlike the standard version, the new protocol frees us from the use of a classical channel to transmit the disambiguation bits, and the use of Pauli's matrices in Bob's side used to reconstruct the teleported state from the mentioned disambiguation bits. These simplifications are the underlying reasons for the title of this paper, that is, simplified protocol.

In the new protocol (Figure 5), block "$|0\rangle$ reset" does a strict reset of the qubit, while, we need to produce $|\beta_{00}\rangle \otimes |\psi\rangle$ instead of $|\psi\rangle \otimes |\beta_{00}\rangle$ used in the standard version. We must highlight as a fundamental contrast between both versions of QTele (the standard and the simplified) that the Kronecker product "$\otimes$" is not commutative [30].

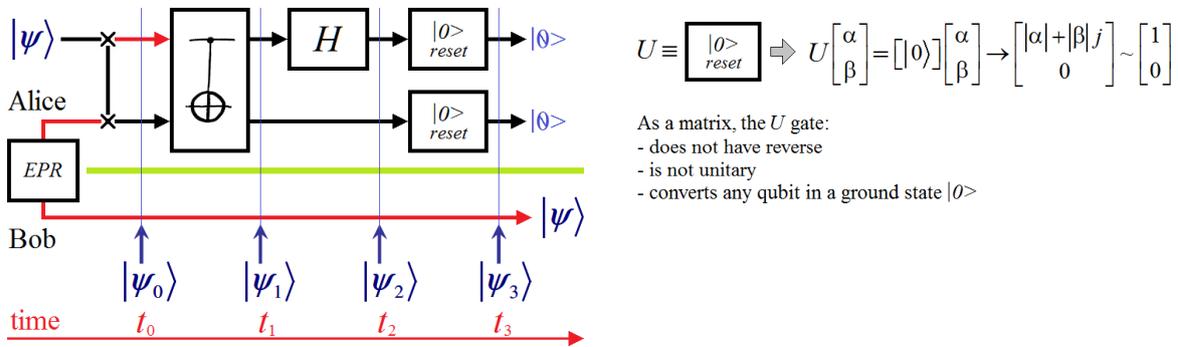

Figure 5. Simplified teleportation protocol using an EPR pair but without classical bits for disambiguation.

$$
\begin{aligned}
|\psi_0\rangle = |\beta_{00}\rangle|\psi\rangle &= \frac{1}{\sqrt{2}}\big(|00\rangle + |11\rangle\big)\big(\alpha|0\rangle + \beta|1\rangle\big) \\
&= \frac{1}{\sqrt{2}}\big[\alpha|000\rangle + \beta|001\rangle + \alpha|110\rangle + \beta|111\rangle\big] \\
&= \Big[\alpha\!\big/\!\sqrt{2} \quad 0 \quad 0 \quad \alpha\!\big/\!\sqrt{2} \quad \beta\!\big/\!\sqrt{2} \quad 0 \quad 0 \quad \beta\!\big/\!\sqrt{2}\Big]^T
\end{aligned}
\tag{17}
$$

Now, a CNOT gate is applied to Equation (17), and the result will be present on the Alice's lower branch, because in her upper branch will be $|\beta_{00}\rangle$,

$$
\begin{aligned}
|\psi_1\rangle &= \frac{1}{\sqrt{2}}\big[\alpha|000\rangle + \beta|001\rangle + \alpha|100\rangle + \beta|101\rangle\big] = \frac{1}{\sqrt{2}}\big[|00\rangle\big(\alpha|0\rangle + \beta|1\rangle\big) + |10\rangle\big(\alpha|0\rangle + \beta|1\rangle\big)\big] \\
&= \frac{1}{\sqrt{2}}\big(\alpha|0\rangle + \beta|1\rangle\big)\big[|00\rangle + |10\rangle\big] = \big(\alpha|0\rangle + \beta|1\rangle\big)\Big[\frac{1}{\sqrt{2}}\big(|0\rangle + |1\rangle\big)\Big]|0\rangle = |\psi\rangle|+\rangle|0\rangle
\end{aligned}
\tag{18}
$$

Figure 5 shows that the Hadamard's gate only involves to the Alice's upper branch, then,

$$
\begin{aligned}
\big(I \otimes H\big)|\beta_{00}\rangle &= \left(\begin{bmatrix} 1 & 0 \\ 0 & 1 \end{bmatrix} \otimes \frac{1}{\sqrt{2}}\begin{bmatrix} 1 & 1 \\ 1 & -1 \end{bmatrix}\right)\frac{1}{\sqrt{2}}\big(|00\rangle + |11\rangle\big) \\
&= \frac{1}{2}\big(|00\rangle + |10\rangle + |01\rangle - |11\rangle\big)
\end{aligned}
\tag{19}
$$

It is clear from Equation (18) that no disambiguation is necessary. Alice blocks her two branch thanks to a qubit reset gate [|0>] pair in order to annul all the projections with |1> components in Eq.(18) and (19). The No-Cloning Theorem [2] is never violated. We can also see in Figure 5 that it is not necessary for Bob to apply any unitary transformation. This eliminates the classical channel that is responsible for making teleportation as a whole to be carried out in a time greater than zero, i.e., not being instantaneous.

Although this result seems to contradict the relativistic principle of causality [4], the reality is that this never happens. As we can see in [25], the instantaneity of entanglement is possible without the need to resort to superluminal signaling and without any contradictions between QMech and SRela. This last fact then covers the new protocol in a direct and complete way.

*2.6 Simplified Quantum Teleportation - noisy analysis*

For noisy EPR pairs we also resorted to Figure 5 using the same version of Equations (12) and (13). Then, repeating Equation (17) but with $|\beta_{00}\rangle_n$ instead of $|\beta_{00}\rangle$, we will have

$$
\begin{aligned}
|\psi_0\rangle = |\beta_{00}\rangle_n |\psi\rangle &= \left(A|00\rangle + B|11\rangle\right)\left(\alpha|0\rangle + \beta|1\rangle\right) \\
&= A\alpha|000\rangle + B\alpha|110\rangle + A\beta|001\rangle + B\beta|111\rangle
\end{aligned}
\tag{20}
$$

Now, we apply a CNOT gate to Equation (20), and the result will be present on the Alice's lower branch again, because in her upper branch will be $|\beta_{00}\rangle$,

$$
\begin{aligned}
|\psi_1\rangle &= A\alpha|000\rangle + B\alpha|100\rangle + A\beta|001\rangle + B\beta|101\rangle \\
&= A|00\rangle\left(\alpha|0\rangle + \beta|1\rangle\right) + B|10\rangle\left(\alpha|0\rangle + \beta|1\rangle\right) \\
&= \left(A|00\rangle + B|10\rangle\right)\left(\alpha|0\rangle + \beta|1\rangle\right) \\
&= \left(A|00\rangle + B|10\rangle\right)|\psi\rangle \\
&= C|\psi\rangle
\end{aligned}
\tag{21}
$$

where,

$$
C = \left(A|00\rangle + B|10\rangle\right)
\tag{22}
$$

In this case again, the Hadamard's gate only involves to the Alice's upper branch, then,

$$
\begin{aligned}
\left(I \otimes H\right)|\beta_{00}\rangle_n &= \left(\begin{bmatrix} 1 & 0 \\ 0 & 1 \end{bmatrix} \otimes \frac{1}{\sqrt{2}}\begin{bmatrix} 1 & 1 \\ 1 & -1 \end{bmatrix}\right)\left(A|00\rangle + B|11\rangle\right) \\
&= \frac{1}{\sqrt{2}}\left(A|00\rangle + A|10\rangle + B|01\rangle - B|11\rangle\right)
\end{aligned}
\tag{23}
$$

The worst consequence of noise in the new protocol is that the teleported state loses its purity, which means, it would not be on Bloch's sphere, in the more general case, given that $C \neq 1$, even so, the teleported state is recovered without problems or disambiguation. This clearly indicates that the new protocol is much more robust (immune to noise) than the standard, for generic qubits as well as for CBS.

## 2.7 Interfaces

From SDC emerges an extraordinary set of interfaces for an efficient relationship between classical bits and CBS, and vice-versa. Figure 6 shows the SDC protocol, which is composed of two well-defined blocks: one which is light blue and the other one pink. The light blue block works as a Cl2Qu interface, while the pink block works as a Qu2Cl interface. All this, of course, is exclusively defined to be used with classical bits, and in consequence, CBS.

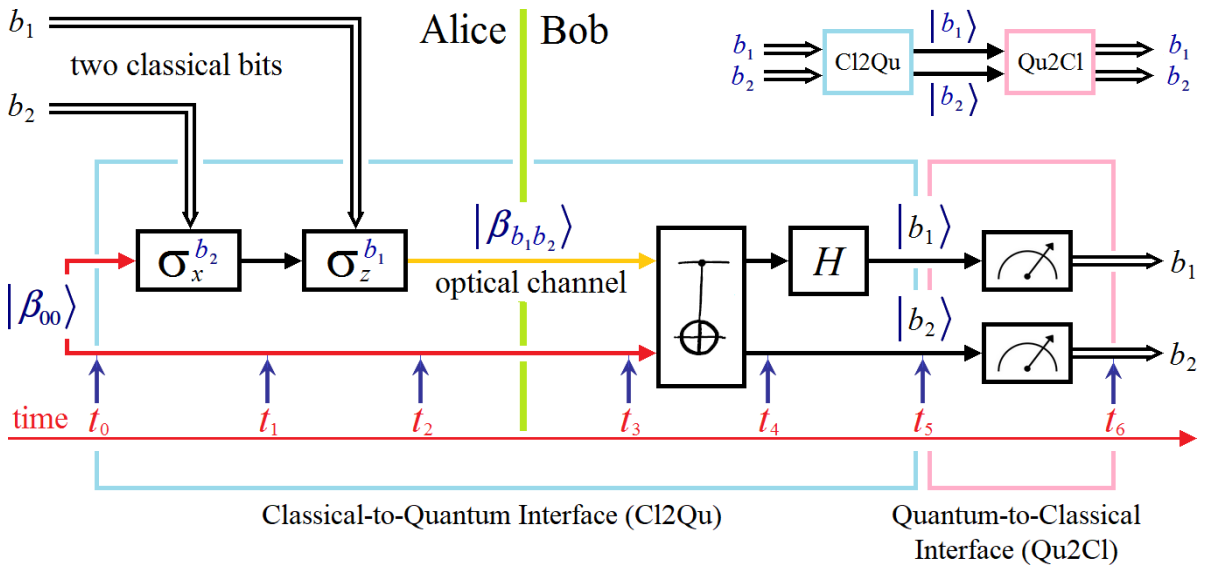

**Figure 6.** SDC protocol and the detail of its constituent interfaces.

Next, we will describe how the SDC protocol works from left to right, which is the correct way of describing how both interfaces (Cl2Qu and Qu2Cl) work. An important detail to highlight before starting is that we can work with: 2 classic bits to be transmitted (Figure 6), a single classic bit ($b_1$) to be transmitted and $b_2$ which makes ancilla equal to zero, or $N$ classic bits to be transmitted as a natural extension of the protocol of Figure 6. This last case will be particularly useful in the practical application of the interfaces for the treatment of digital images, in such a way that $N = 24$ and each application of the protocol implies working with a complete pixel of a color image. For a better development of the idea, we will work with 2 classic bits extracting Cl2Qu from Figure 6 to form Figure 7.

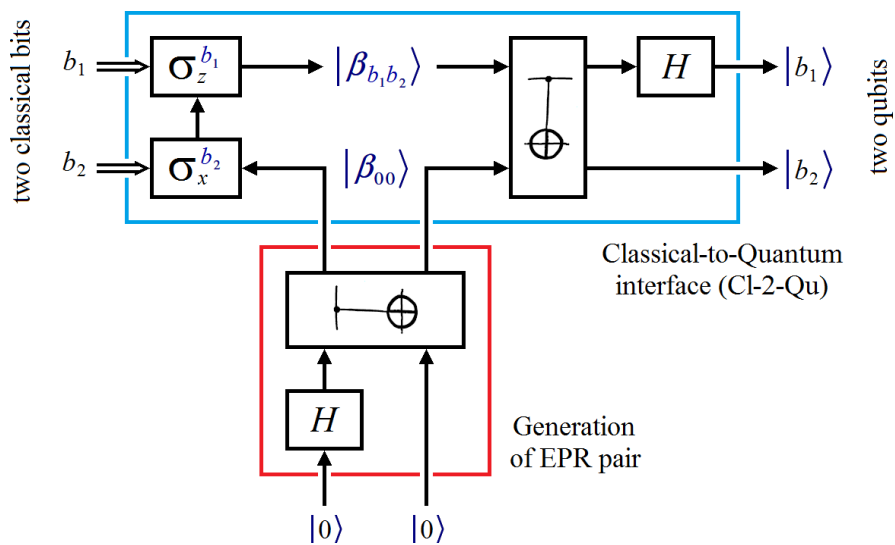

**Figure 7.** Classical-to-Quantum interface.

We begin with both possibilities according to $b_2$, i.e., 0 or 1. If $b_2 = 0$, then, in $t_1$ we will have $|\beta_{00}\rangle$ again (Figure 6). But, if $b_2 = 1$, then, in $t_1$ we will have,

$$\left(\sigma_x \otimes I\right)|\beta_{00}\rangle = \left(\begin{bmatrix} 0 & 1 \\ 1 & 0 \end{bmatrix} \otimes \begin{bmatrix} 1 & 0 \\ 0 & 1 \end{bmatrix}\right)\begin{bmatrix} \frac{1}{\sqrt{2}} \\ 0 \\ 0 \\ \frac{1}{\sqrt{2}} \end{bmatrix} = \begin{bmatrix} 0 & 0 & 1 & 0 \\ 0 & 0 & 0 & 1 \\ 1 & 0 & 0 & 0 \\ 0 & 1 & 0 & 0 \end{bmatrix}\begin{bmatrix} \frac{1}{\sqrt{2}} \\ 0 \\ 0 \\ \frac{1}{\sqrt{2}} \end{bmatrix} = \begin{bmatrix} 0 \\ \frac{1}{\sqrt{2}} \\ \frac{1}{\sqrt{2}} \\ 0 \end{bmatrix} \tag{24}$$

If in $t_2$, $b_2 = 0$ and $b_1 = 0$, we will obtain $|\beta_{00}\rangle$. But if $b_2 = 0$ and $b_1 = 1$, we will have,

$$\left(\sigma_z \otimes I\right)|\beta_{00}\rangle = \left(\begin{bmatrix} 1 & 0 \\ 0 & -1 \end{bmatrix} \otimes \begin{bmatrix} 1 & 0 \\ 0 & 1 \end{bmatrix}\right)\begin{bmatrix} \frac{1}{\sqrt{2}} \\ 0 \\ 0 \\ \frac{1}{\sqrt{2}} \end{bmatrix} = \begin{bmatrix} 1 & 0 & 0 & 0 \\ 0 & 1 & 0 & 0 \\ 0 & 0 & -1 & 0 \\ 0 & 0 & 0 & -1 \end{bmatrix}\begin{bmatrix} \frac{1}{\sqrt{2}} \\ 0 \\ 0 \\ \frac{1}{\sqrt{2}} \end{bmatrix} = \begin{bmatrix} \frac{1}{\sqrt{2}} \\ 0 \\ 0 \\ -\frac{1}{\sqrt{2}} \end{bmatrix} \tag{25}$$

Now, if $b_2 = 1$ and $b_1 = 1$, we will use the result of Equation (24) to obtain the state in $t_2$,

$$\left(\sigma_z \otimes I\right)\begin{bmatrix} 0 \\ \frac{1}{\sqrt{2}} \\ \frac{1}{\sqrt{2}} \\ 0 \end{bmatrix} = \left(\begin{bmatrix} 1 & 0 \\ 0 & -1 \end{bmatrix} \otimes \begin{bmatrix} 1 & 0 \\ 0 & 1 \end{bmatrix}\right)\begin{bmatrix} 0 \\ \frac{1}{\sqrt{2}} \\ \frac{1}{\sqrt{2}} \\ 0 \end{bmatrix} = \begin{bmatrix} 1 & 0 & 0 & 0 \\ 0 & 1 & 0 & 0 \\ 0 & 0 & -1 & 0 \\ 0 & 0 & 0 & -1 \end{bmatrix}\begin{bmatrix} 0 \\ \frac{1}{\sqrt{2}} \\ \frac{1}{\sqrt{2}} \\ 0 \end{bmatrix} = \begin{bmatrix} 0 \\ \frac{1}{\sqrt{2}} \\ -\frac{1}{\sqrt{2}} \\ 0 \end{bmatrix} \tag{26}$$

Obviously, the result in $t_2$ is the same as in $t_3$ (Figure 6). The four obtained results, according to the values of $b_2$ and $b_1$, travel through the optical channel of Figure 6 under the generic name of $|\beta_{b_1 b_2}\rangle$.

Now, if we apply *CNOT* gate to these results, then, for $b_2 = 0$ and $b_1 = 0$, we will have,

$$\begin{bmatrix} 1 & 0 & 0 & 0 \\ 0 & 1 & 0 & 0 \\ 0 & 0 & 0 & 1 \\ 0 & 0 & 1 & 0 \end{bmatrix}\begin{bmatrix} \frac{1}{\sqrt{2}} \\ 0 \\ 0 \\ \frac{1}{\sqrt{2}} \end{bmatrix} = \begin{bmatrix} \frac{1}{\sqrt{2}} \\ 0 \\ \frac{1}{\sqrt{2}} \\ 0 \end{bmatrix} \tag{27}$$

For $b_2 = 0$ and $b_1 = 1$,

$$\begin{bmatrix} 1 & 0 & 0 & 0 \\ 0 & 1 & 0 & 0 \\ 0 & 0 & 0 & 1 \\ 0 & 0 & 1 & 0 \end{bmatrix}\begin{bmatrix} \frac{1}{\sqrt{2}} \\ 0 \\ 0 \\ -\frac{1}{\sqrt{2}} \end{bmatrix} = \begin{bmatrix} \frac{1}{\sqrt{2}} \\ 0 \\ -\frac{1}{\sqrt{2}} \\ 0 \end{bmatrix} \tag{28}$$

For $b_2 = 1$ and $b_1 = 0$,

$$\begin{bmatrix} 1 & 0 & 0 & 0 \\ 0 & 1 & 0 & 0 \\ 0 & 0 & 0 & 1 \\ 0 & 0 & 1 & 0 \end{bmatrix}\begin{bmatrix} 0 \\ 1/\sqrt{2} \\ 1/\sqrt{2} \\ 0 \end{bmatrix} = \begin{bmatrix} 0 \\ 1/\sqrt{2} \\ 0 \\ 1/\sqrt{2} \end{bmatrix} \tag{29}$$

For $b_2 = 1$ and $b_1 = 1$,

$$\begin{bmatrix} 1 & 0 & 0 & 0 \\ 0 & 1 & 0 & 0 \\ 0 & 0 & 0 & 1 \\ 0 & 0 & 1 & 0 \end{bmatrix}\begin{bmatrix} 0 \\ 1/\sqrt{2} \\ -1/\sqrt{2} \\ 0 \end{bmatrix} = \begin{bmatrix} 0 \\ 1/\sqrt{2} \\ 0 \\ -1/\sqrt{2} \end{bmatrix} \tag{30}$$

At this point, it is important to clarify that the quantum channel of Figure 6 inside the Cl2Qu interface is obviously unnecessary (Figure 7). In fact, the definitive Cl2Qu interface is absolutely compact and henceforth it will be considered as a unified block. Finally, we will apply the Hadamard's gate to the last set of equations according to Figures 6 and 7 [24, 25], thus, for $b_2 = 0$ and $b_1 = 0$, we will have,

$$\begin{aligned} \left(H \otimes I\right)\begin{bmatrix} 1/\sqrt{2} \\ 0 \\ 1/\sqrt{2} \\ 0 \end{bmatrix} &= \left(\begin{bmatrix} 1/\sqrt{2} & 1/\sqrt{2} \\ 1/\sqrt{2} & -1/\sqrt{2} \end{bmatrix} \otimes \begin{bmatrix} 1 & 0 \\ 0 & 1 \end{bmatrix}\right)\begin{bmatrix} 1/\sqrt{2} \\ 0 \\ 1/\sqrt{2} \\ 0 \end{bmatrix} = \begin{bmatrix} 1/\sqrt{2} & 0 & 1/\sqrt{2} & 0 \\ 0 & 1/\sqrt{2} & 0 & 1/\sqrt{2} \\ 1/\sqrt{2} & 0 & -1/\sqrt{2} & 0 \\ 0 & 1/\sqrt{2} & 0 & -1/\sqrt{2} \end{bmatrix}\begin{bmatrix} 1/\sqrt{2} \\ 0 \\ 1/\sqrt{2} \\ 0 \end{bmatrix} \\ &= \begin{bmatrix} 1 \\ 0 \\ 0 \\ 0 \end{bmatrix} = \begin{bmatrix} 1 \\ 0 \end{bmatrix} \otimes \begin{bmatrix} 1 \\ 0 \end{bmatrix} = |0\rangle \otimes |0\rangle = |00\rangle \end{aligned} \tag{31}$$

for $b_2 = 0$ and $b_1 = 1$,

$$\begin{aligned} \left(H \otimes I\right)\begin{bmatrix} 1/\sqrt{2} \\ 0 \\ -1/\sqrt{2} \\ 0 \end{bmatrix} &= \left(\begin{bmatrix} 1/\sqrt{2} & 1/\sqrt{2} \\ 1/\sqrt{2} & -1/\sqrt{2} \end{bmatrix} \otimes \begin{bmatrix} 1 & 0 \\ 0 & 1 \end{bmatrix}\right)\begin{bmatrix} 1/\sqrt{2} \\ 0 \\ -1/\sqrt{2} \\ 0 \end{bmatrix} = \begin{bmatrix} 1/\sqrt{2} & 0 & 1/\sqrt{2} & 0 \\ 0 & 1/\sqrt{2} & 0 & 1/\sqrt{2} \\ 1/\sqrt{2} & 0 & -1/\sqrt{2} & 0 \\ 0 & 1/\sqrt{2} & 0 & -1/\sqrt{2} \end{bmatrix}\begin{bmatrix} 1/\sqrt{2} \\ 0 \\ -1/\sqrt{2} \\ 0 \end{bmatrix} \\ &= \begin{bmatrix} 0 \\ 0 \\ 1 \\ 0 \end{bmatrix} = \begin{bmatrix} 0 \\ 1 \end{bmatrix} \otimes \begin{bmatrix} 1 \\ 0 \end{bmatrix} = |1\rangle \otimes |0\rangle = |10\rangle \end{aligned} \tag{32}$$

for $b_2 = 1$ and $b_1 = 0$,

$$\left(H\otimes I\right)\begin{bmatrix}0\\ 1/\sqrt{2}\\ 0\\ 1/\sqrt{2}\end{bmatrix}=\left(\begin{bmatrix}1/\sqrt{2}&1/\sqrt{2}\\ 1/\sqrt{2}&-1/\sqrt{2}\end{bmatrix}\otimes\begin{bmatrix}1&0\\ 0&1\end{bmatrix}\right)\begin{bmatrix}0\\ 1/\sqrt{2}\\ 0\\ 1/\sqrt{2}\end{bmatrix}=\begin{bmatrix}1/\sqrt{2}&0&1/\sqrt{2}&0\\ 0&1/\sqrt{2}&0&1/\sqrt{2}\\ 1/\sqrt{2}&0&-1/\sqrt{2}&0\\ 0&1/\sqrt{2}&0&-1/\sqrt{2}\end{bmatrix}\begin{bmatrix}0\\ 1/\sqrt{2}\\ 0\\ 1/\sqrt{2}\end{bmatrix}$$

$$=\begin{bmatrix}0\\1\\0\\0\end{bmatrix}=\begin{bmatrix}1\\0\end{bmatrix}\otimes\begin{bmatrix}0\\1\end{bmatrix}=\left|0\right\rangle\otimes\left|1\right\rangle=\left|01\right\rangle \tag{33}$$

for $b_2 = 1$ and $b_1 = 1$,

$$\left(H\otimes I\right)\begin{bmatrix}0\\ 1/\sqrt{2}\\ 0\\ -1/\sqrt{2}\end{bmatrix}=\left(\begin{bmatrix}1/\sqrt{2}&1/\sqrt{2}\\ 1/\sqrt{2}&-1/\sqrt{2}\end{bmatrix}\otimes\begin{bmatrix}1&0\\ 0&1\end{bmatrix}\right)\begin{bmatrix}0\\ 1/\sqrt{2}\\ 0\\ -1/\sqrt{2}\end{bmatrix}=\begin{bmatrix}1/\sqrt{2}&0&1/\sqrt{2}&0\\ 0&1/\sqrt{2}&0&1/\sqrt{2}\\ 1/\sqrt{2}&0&-1/\sqrt{2}&0\\ 0&1/\sqrt{2}&0&-1/\sqrt{2}\end{bmatrix}\begin{bmatrix}0\\ 1/\sqrt{2}\\ 0\\ -1/\sqrt{2}\end{bmatrix}$$

$$=\begin{bmatrix}0\\0\\0\\1\end{bmatrix}=\begin{bmatrix}0\\1\end{bmatrix}\otimes\begin{bmatrix}0\\1\end{bmatrix}=\left|1\right\rangle\otimes\left|1\right\rangle=\left|11\right\rangle \tag{34}$$

Here Cl2Qu interface finalizes. Basically, Cl2Qu interface is a block that performs a transfer of type $\{b_1, b_2\} \to \{|b_1\rangle, |b_2\rangle\}$ for CBS, i.e., $\{0,1\} \to \{|0\rangle, |1\rangle\}$.

The pink block of Figure 6 constitutes the Qu2Cl interface. In fact, at the entrance of that block, we have $\{|b_1\rangle, |b_2\rangle\}$, while at its exit we have $\{b_1, b_2\}$, i.e., QuMe is a Qu2Cl interface in itself. With this sequence Cl2Qu + Qu2Cl we complete the standard SDC. In this way, the circle closes, and at least in theory, we should recover at the exit of the pink block the same classical bits that we have entered at the entrance of the light blue block. Consequently, we can use a coincidence counter to evaluate the performance of the complete SDC protocol, and to know what the level of coupling between both interfaces is.

A relevant detail to take into account is that when we measure a CBS, we completely recover its classical counterpart, a situation very different from what happens with a generic qubit, which does not have a classical counterpart and if it existed it would be unattainable to obtain it [30, 31, 33]. In fact, it is only necessary to make a measurement on the *z*-axis of the Bloch's sphere (vertical) avoiding the typical problems associated with QuMe, HUP and reciprocity [30-33].

### 2.8 Complete architecture to be used

Figure 8 represents a complete architecture for an enhanced SDC. This figure shows both interfaces where we can appreciate a pair of green blocks labeled QTele between them. Actually, these blocks can represent the two types of seen teleportations: standard and simplified. Besides, these blocks constitute the imaginary boundary between Alice's and Bob's sides. It is important to mention that this configuration can be used to transmit a single classic bit at a time, where, $b_2$ could be considered an ancilla, then, the architecture of Figure 8 would have a single block of QTele or horizontal thread.

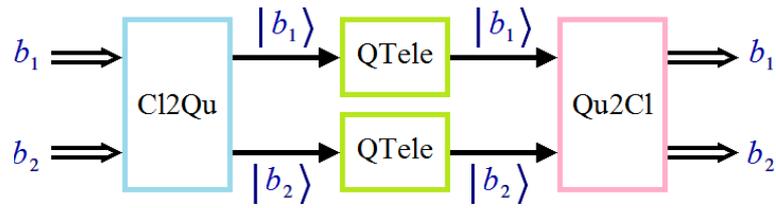

**Figure 8.** Enhanced SDC using Cl2Qu and Qu2Cl interfaces and two blocks of QTele.

If we expand the idea of Figure 8 by involving the steps necessary to decompose a color digital image in their corresponding constituent bits, then, we will arrive to Figure 9. This is essentially the scheme we will use for the experiments in the next section. Figure 9 shows on top the original image to be teleported, then the three color channels (red, green and blue) of the image are separated. Later, each color channel is decomposed in 8 bitplanes. The bit of each bitplane is introduced into a Cl2Qu interface. The equivalent qubits are teleported. Bob receives each qubit with a Qu2Cl interface. With the obtained bits, the bitplanes are reconstructed, then the three color channels are reassembled and finally the teleported image is obtained. The lower part of Figure 9 shows a comparison between the acquired bit of the original image and the recovered bit from the complete architecture. Finally, the complete comparison of all individual bits becomes the total comparison between both images: the original and the teleported one for the purpose of evaluating the level of degradation introduced by the complete procedure using as a metric the coincidence counter. Notwithstanding the foregoing, in the following section, the intermediate instances of each step of the procedure will also be monitored and analyzed.

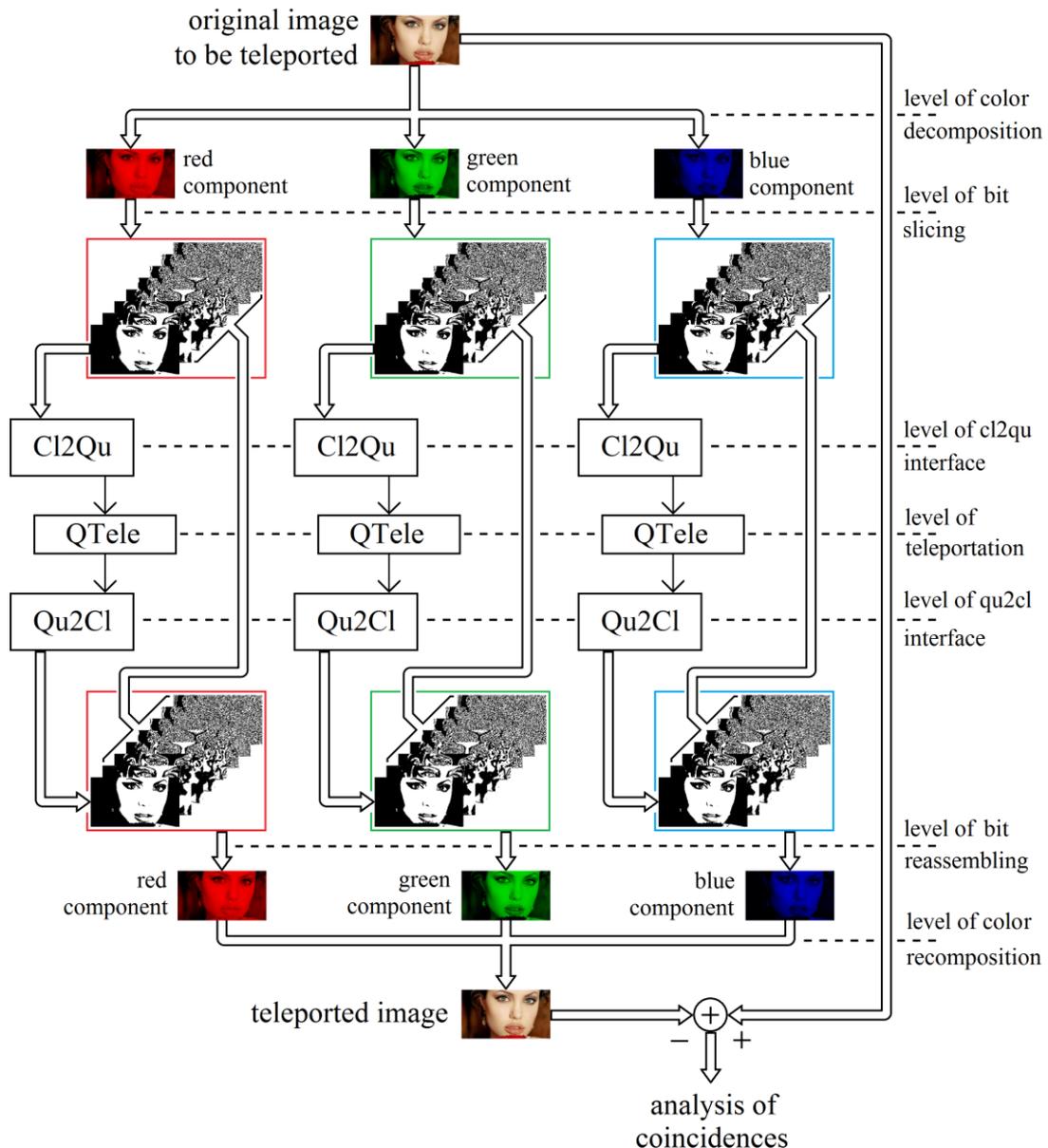

**Figure 9.** Complete teleportation of a 3-color-digital-image (*Angelina*).

## 3. Results

### 3.1. Setup

All experiments of this section consist of implementations on Quirk® [44] simulator. In the complete architecture for the QTele of *Angelina* image (Figure 9), we randomly select 100 of the 1920x1080x3x8 bits of the original image, although fewer bits of alternate values (i.e., 0 and 1) and from different locations would be enough.

We will evaluate the performance in the reconstruction of such bits thanks to a coincidence counter, which can be seen in the lower part of Figure 9.

Finally, we will clarify the number of qubits used in each simulation carried out with the Quirk® simulator, as well as the corresponding circuit lay-out, in order to facilitate the reproduction of all the experiments done here.

### 3.2. Partial tests

These experiments involve evaluating separately each of the protocols to be used in the final configuration for the QTele of the image, before proceeding with it. Such experiments aim to unmask in an individual way the possible responsible for the collectively incorrect results.

#### 3.2.1. Superdense coding

Based on the protocol of Figure 6, we will implement the complete SDC, which is equivalent to the union between Cl2Qu and Qu2Cl interfaces. Figure 10 represents the complete configuration for Quirk® simulator with the explicit results inside the figure through a series of activation (on) and deactivation (off) flags, from $\{b_1,b_2\}=\{0,0\}$ to $\{b_1,b_2\}=\{1,1\}$.

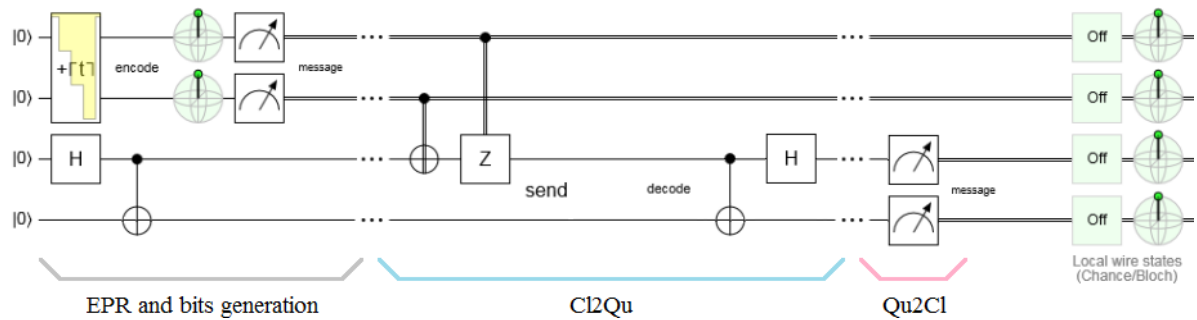

**Figure 10(a)**. SDC ≡ Cl2Qu ∪ Qu2Cl on Quirk® for $\{b_1,b_2\}=\{0,0\}$.

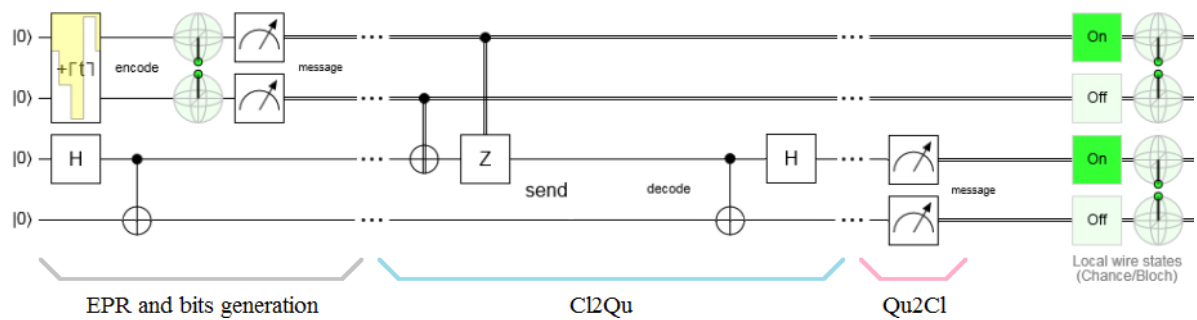

**Figure 10(b)**. SDC ≡ Cl2Qu ∪ Qu2Cl on Quirk® for $\{b_1,b_2\}=\{1,0\}$.

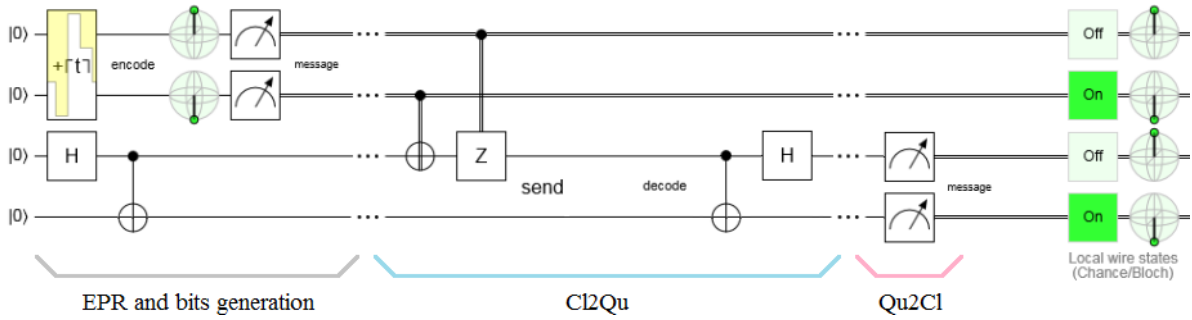

**Figure 10(c)**. SDC ≡ Cl2Qu ∪ Qu2Cl on Quirk® for {$b_1$,$b_2$}={0,1}.

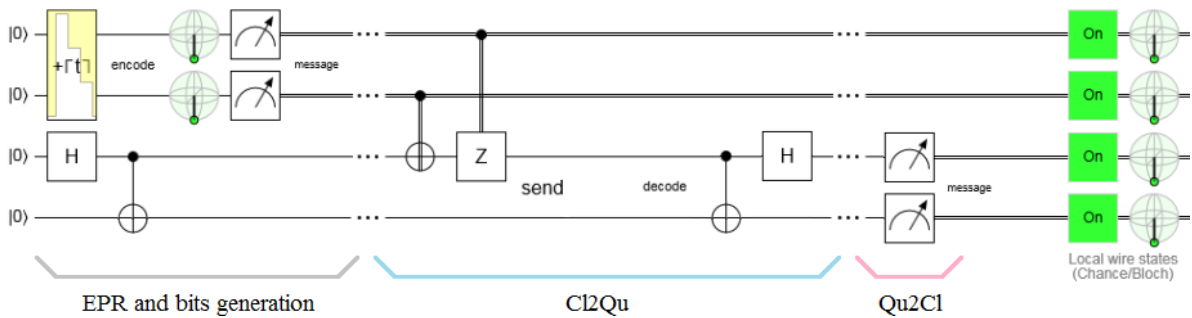

**Figure 10(d)**. SDC ≡ Cl2Qu ∪ Qu2Cl on Quirk® for {$b_1$,$b_2$}={1,1}.

For this experiment, we used 4 qubits on the Quirk® simulator. The total coincidences of the flags (in on or off, at the entrance and exit of the protocol) in the four cases (i.e., 00, 01, 10 and 11) represent the success in the SDC process. The connection labelled as *send* represents the optical channel in yellow in Figure 6, although, as it is obvious, within the Cl2Qu interface such a connection does not exist. Finally, 3 points in a row represent the border between both blocks.

The experimental implementation for the transmission of 100 classical bits taken into pairs and their posterior recovering gave us as a result on Quirk® simulator a complete set of coincidences. It is evident that fewer pairs would have given the same results.

### 3.2.2. Standard QTele

Figure 11 shows the complete process of Standard QTele, which is only implemented thanks to 3 qubits on Quirk® simulator.

Figure 11(a) remits us to the case of a generic qubit, which has an arbitrary allocation on the Bloch's sphere [30]. Figures 11(b) and 11(c) show a pair of particular cases where the qubits are both CBS, i.e., $|0\rangle$ and $|1\rangle$, respectively.

As in the case of Figures 4 and 5, a single fine line represents a wire carrying one qubit, while a double line represents a wire carrying one classical bit [30].

The possibility of a fast, precise and graphically expressive simulation on a complete toolbox environment, independently of the pertinent code makes Quirk® simulator an attractive choice.

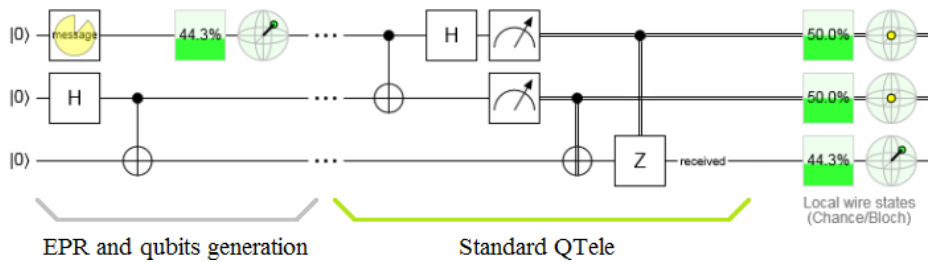

**Figure 11(a)**. Standard QTele for a generic qubit to be teleported.

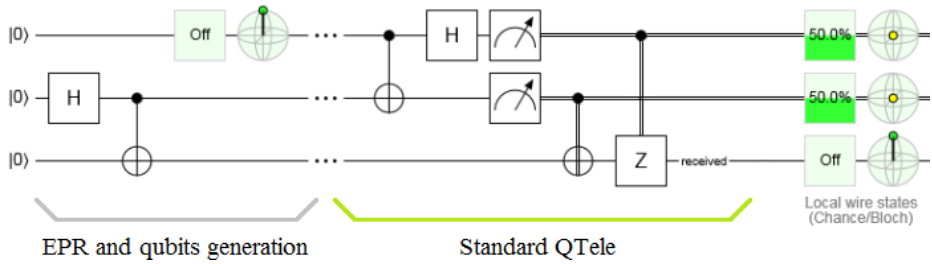

**Figure 11(b)**. Standard QTele for the teleportation of $|0\rangle$.

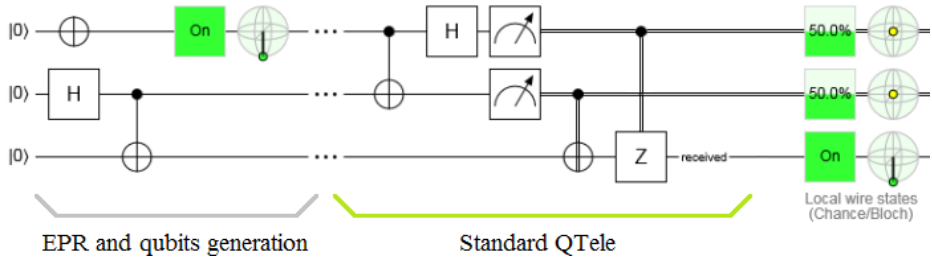

**Figure 11(c)**. Standard QTele for the teleportation of $|1\rangle$.

### 3.2.3. Simplified QTele

Figure 12 shows the Simplified QTele with identical considerations to the previous case.

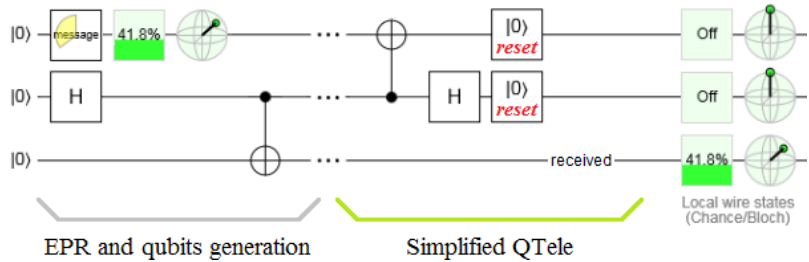

**Figure 12(a)**. Simplified QTele for a generic qubit to be teleported.

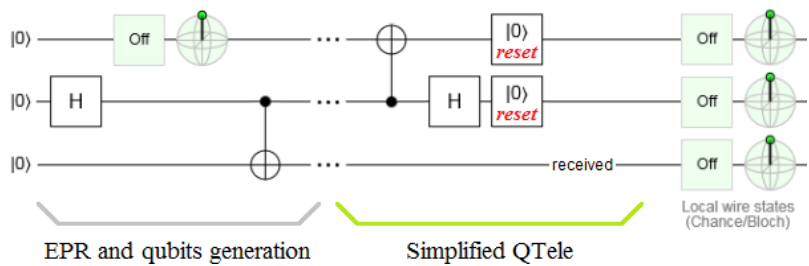

**Figure 12(b)**. Simplified QTele for the teleportation of $|0\rangle$.

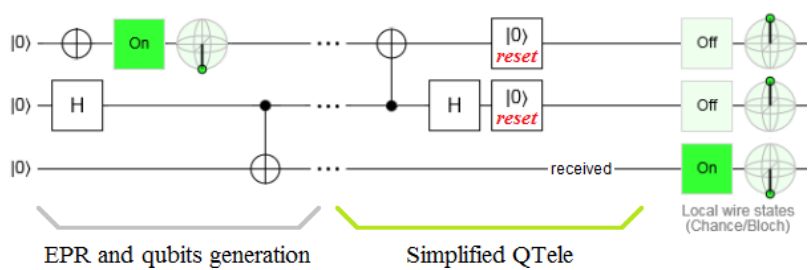

**Figure 12(c)**. Simplified QTele for the teleportation of $|1\rangle$.

### 3.3. Standard QTele of a digital image

As we have said, this is the first implementation for the QTele of an image. Figure 13 represents such implementation on Quirk® simulator.

The architecture of Figure 13 shows the perfect complementation between the Standard QTele and both interfaces, i.e., Cl2Qu and Qu2Cl. This configuration required 8 qubits.

We must bear in mind that at a certain point the QTele must be separated a considerable distance between Alice's side and Bob's side. In Figure 13 that separation has not been incorporated so as not to complicate the graphic.

The 50 pairs of classic bits of type {00,01,10,11} were randomly taken from all those that makeup *Angelina* image, in fact, 1920x1080x3x8, according to the procedure of Figure 9. The percentage of coincidences reached 100 %.

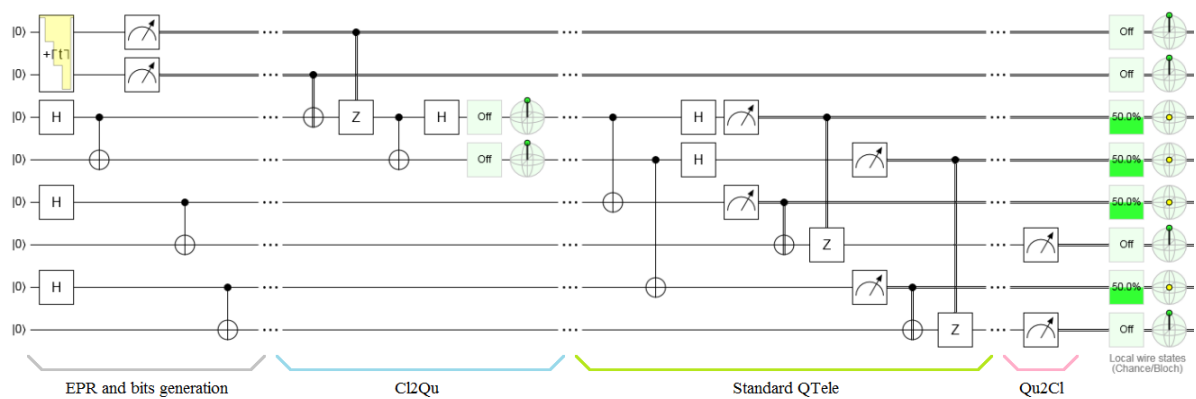

**Figure 13(a)**. Standard QTele on Quirk® of a pair {$b_1$,$b_2$}={0,0}.

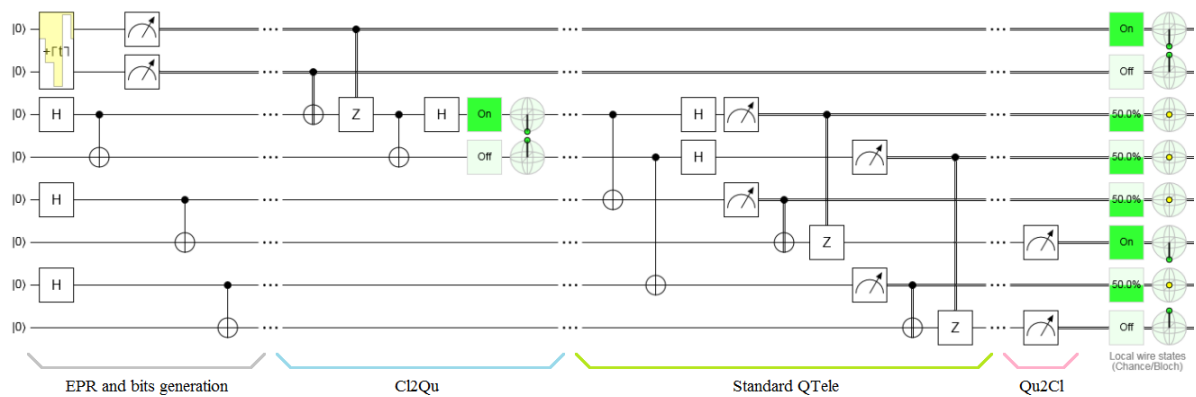

**Figure 13(b)**. Standard QTele on Quirk® of a pair {$b_1$,$b_2$}={0,1}.

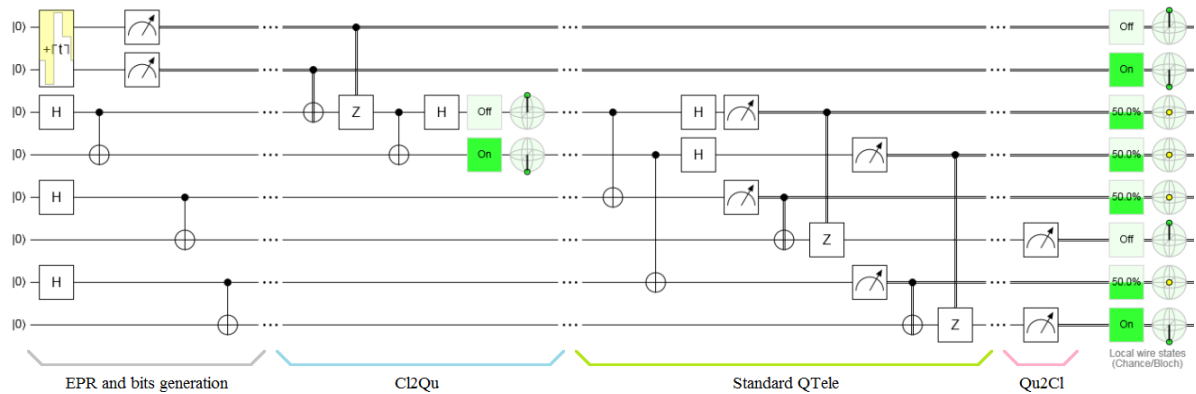

**Figure 13(c)**. Standard QTele on Quirk® of a pair {$b_1$,$b_2$}={1,0}.

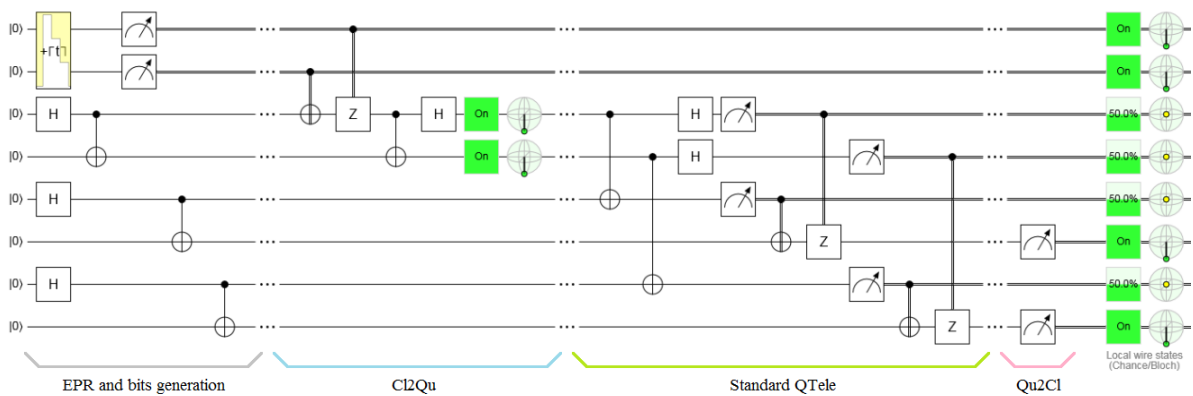

**Figure 13(d)**. Standard QTele on Quirk® of a pair {$b_1$,$b_2$}={1,1}.

### 3.4. Simplified QTele of a digital image

Figure 14 shows us the economy in the gates utilization of the Simplified version respect to the implementation based on the Standard version of QTele, however, we have obtained similar results in both cases with 8 qubits too.

For this case, we also used 50 pairs of bits randomly selected from the complete set of bits corresponding to the original image of *Angelina*. Figure 14 represents the four cases of possible pairs that we found during the random exploration of the image. These pairs are the same that we used in the previous case and for SDC, i.e., {00,01,10,11}.

This implementation, like the previous ones, demonstrates the effectiveness and ductility of the Cl2Qu and Qu2Cl interfaces, which can be successfully reused in other areas of QIP [30], such as QComp [45], QComm [26-29], and Quantum Internet [46-49], including undoubtedly, Quantum Cryptography [50], in general, and Quantum Key Distribution (QKD) [51-54], in particular.

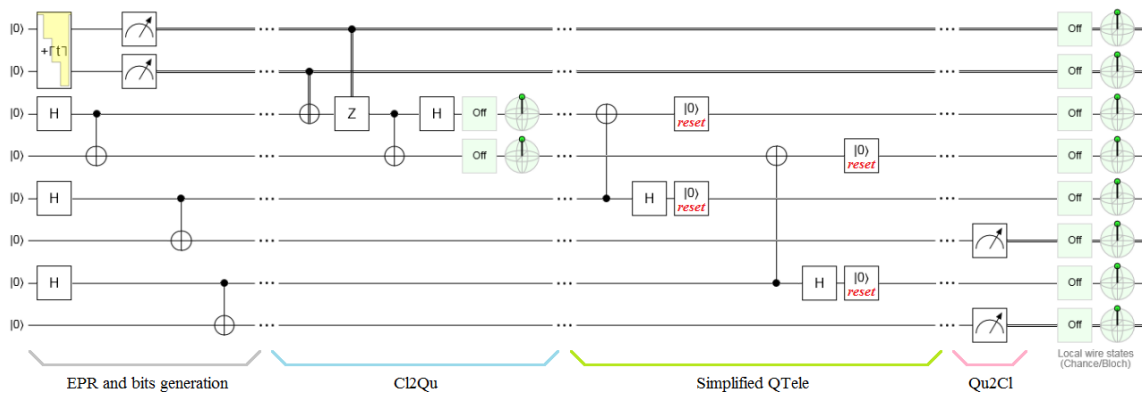

**Figure 14(a)**. Simplified QTele on Quirk® of a pair {$b_1$,$b_2$}={0,0}.

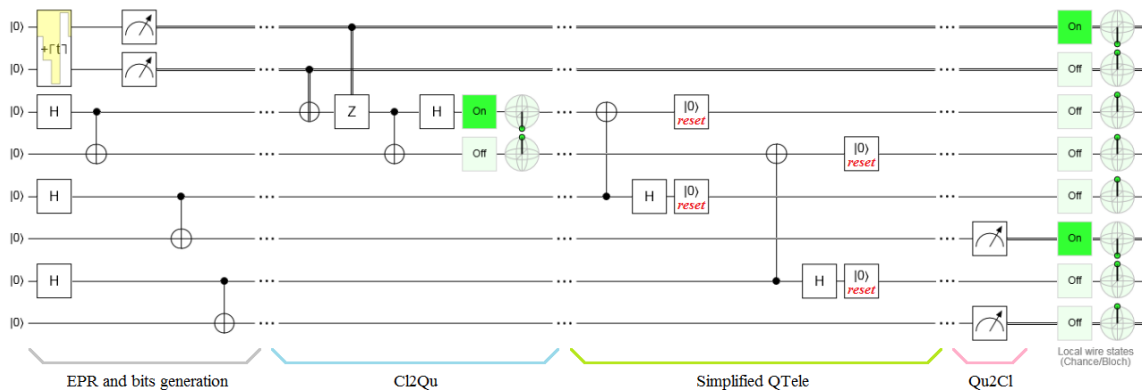

**Figure 14(b)**. Simplified QTele on Quirk® of a pair {$b_1$,$b_2$}={0,1}.

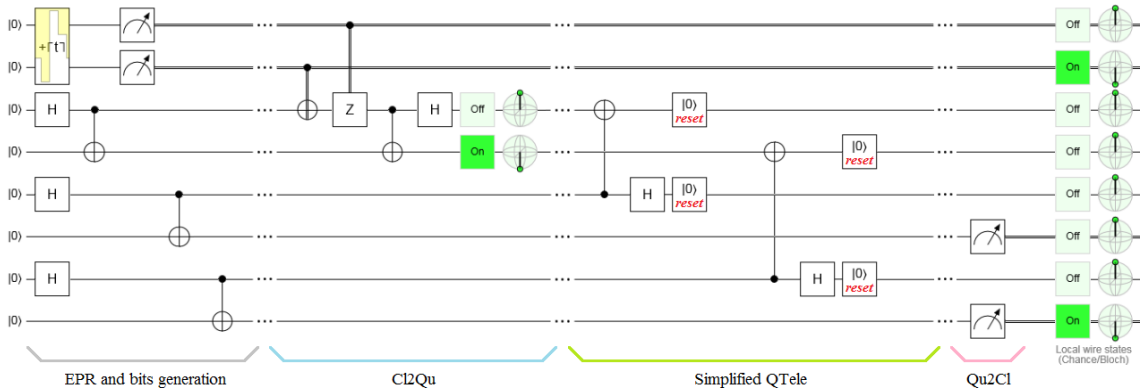

**Figure 14(c)**. Simplified QTele on Quirk® of a pair {$b_1,b_2$}={1,0}.

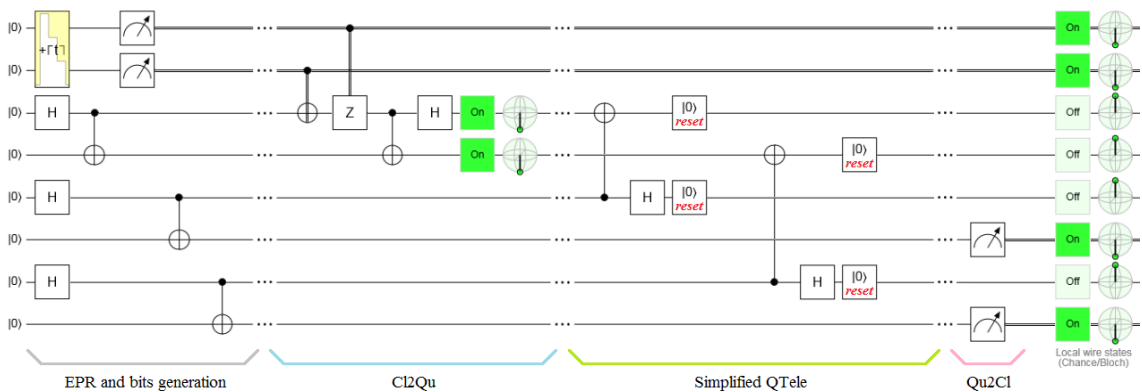

**Figure 14(d)**. Simplified QTele on Quirk® of a pair {$b_1,b_2$}={1,1}.

## 4. Discussion

The results of Section 3 show the absolute viability of the procedure consisting of the QTele of an image, as well as, the complete success of the new interfaces: Cl2Qu and Qu2Cl. Besides, the simplified version of QTele has demonstrated to have similar results to the standard version, which is plausible considering the fact that it has fewer quantum gates than the standard version. The two QTele versions were successful in both the partial and the collective experiments.

The fact that both integrating experiments begin and end with the binary bits belonging to the pixels of the image to be teleported and the teleported image, respectively, allowed us the use of a coincidence counter as a metric. This constituted an excellent strategy for evaluating the functional quality of the integrating architecture, with the same strictness than those individually used in both QTele protocols and both interfaces.

Another important strategy consisted of the use of Quirk® simulator in each experiment. It let us know the final outcome following a binary criterion: *it works-or-it does not work*. This is of great value since knowing the final outcome, whether or not a specific configuration or architecture works, is key before mobilizing a huge amount of human resources as well as purchasing and/or replacing expensive laboratory equipment for an eventual optical table.

The findings have deep implications in the context of all type of digital signal transmission, where, such signals can represent streaming, multi and hyper-spectral images, video or future TV broadcasting. Besides, the new interfaces have an excellent projection on QIP [30] in general and Quantum Computing [45] in particular, in configurations like Cl2Qu-*Quantum-Algorithm*-Qu2Cl.

Future implementations will be directed to applications outside the laboratory, i.e., practical uses like Quantum Internet [46-49]. Besides, the projection of the simplified QTele on Quantum Key Distribution (QKD) [51-54] is very interesting, since this new protocol does not use classical bits for disambiguation via a classical channel, while increasing considerably the level of security of QKD eliminating well an unnecessary exposition on one of the channels commonly used in it.

On the other hand, we must bear in mind that for both integrating architectures we must distribute 1920x1080x3x8/2 EPR pairs, in other words, a logistical nightmare. Besides, we must consider the following:

- the short half-life of the entanglement due to decoherence [55],
- the short half-life of the qubits [56],
- the type of used qubit (trapped ion, superconductor, topological, etc.) [57], and
- the way of distributing the EPR pairs: delivery vs take-out [25].

Finally, Figures 10 to 14 were edited specifically to incorporate the lower labels that demarcate each constituent block only for a better understanding of the original Quirk® simulator outcome.

**Conflicts of Interest:** The author declares that there are no competing interests.

**Funding Statement:** The author acknowledges funding by LosWW under contract QComm-06#05/01/2015.

ACKNOWLEDGMENTS: M. Mastriani thanks boarding of LosWW for his tremendous help and support.